\documentclass{article}
\usepackage[utf8]{inputenc}

\PassOptionsToPackage{hyphens}{url}
\usepackage[bookmarks=false,hidelinks]{hyperref}

\usepackage{booktabs}
\usepackage{spconf,amsmath,graphicx}
\usepackage{multirow,balance,cite}

\title{I4U System Description FOR NIST SRE'20 CTS Challenge}

\name{\parbox{\linewidth}{\centering
Kong Aik Lee$^{1}$, Tomi Kinnunen$^{2}$, Daniele Colibro$^{3}$, Claudio Vair$^{3}$, Andreas Nautsch$^{4}$, Hanwu Sun$^{1}$,\\ Liang He$^{5}$, Tianyu Liang$^{5}$, Qiongqiong Wang$^{6}$,  Mickael Rouvier$^{7}$, Pierre-Michel Bousquet$^{7}$,\\ Rohan Kumar Das$^{8}$, Ignacio Viñals Bailo$^{9}$, Meng Liu$^{10}$, Héctor Deldago$^{11}$,\\ Xuechen Liu$^{2,12}$, Md Sahidullah$^{12}$, Sandro Cumani$^{3}$, Boning Zhang$^{5}$, Koji Okabe$^{6}$, \\ Hitoshi Yamamoto$^{6}$, Ruijie Tao$^{8}$, Haizhou Li$^{8}$, Alfonso Ortega Giménez$^{9}$,\\ Longbiao Wang$^{10}$, Luis Buera$^{11}$}}

\address{
  $^{1}$Institute for Infocomm Research, A*STAR, Singapore\\
  $^{2}$School of Computing, University of Eastern Finland, Finland\\
  $^{3}$Loquendo and Politecnico di Torino, Italy\\
  $^{4}$EURECOM, France and vitas.ai, Germany\\
  $^{5}$Department of Electronic Engineering, Tsinghua University, China\\
  $^{6}$NEC Corporation, Japan\\
  $^{7}$LIA, Avignon University, France\\
  $^{8}$National University of Singapore, Singapore\\
  $^{9}$Vivolab, Spain\\
  $^{10}$Tianjian University, China\\
  $^{11}$Nuance Communications Inc., Spain\\
  $^{12}$INRIA, France
  }

\begin{document}

\maketitle

\begin{abstract}
This manuscript describes the I4U submission to the 2020 NIST Speaker Recognition Evaluation (SRE'20) Conversational Telephone Speech (CTS) Challenge. The I4U's submission was resulted from active collaboration among researchers across eight research teams -- I$^2$R (Singapore), UEF (Finland), VALPT (Italy, Spain), NEC (Japan), THUEE (China), LIA (France), NUS (Singapore), INRIA (France) and TJU (China). The submission was based on the fusion of top performing sub-systems and sub-fusion systems contributed by individual teams. Efforts have been spent on the use of common development and validation sets, submission schedule and milestone, minimizing inconsistency in trial list and score file format across sites. 

\end{abstract}

\section{Introduction}

The I4U submission is a joint effort of cross-site research teams - I$^2$R (Singapore), UEF (Finland), Loquendo and Politecnico di Torino (Italy), VivoLab and Agnitio (Spain), NEC (Japan), THUEE (China), LIA (France), NUS (Singapore), INRIA (France) and TJU (China). The SRE'20 CTS Challenge is an extension to the SRE'19 CTS task with an additional challenge being multi-lingual evaluation set. We paid significant efforts in keeping a common training and development set, submission schedule and milestone, minimizing inconsistency in trial lists across sites. This manuscript presents the technical details of the datasets, sub-systems development, and fusion system.

We adopted a fusion strategy where sub-systems are fused with sub-fusion systems. These sub-systems are based on latest advances in neural speaker embeddings, namely, x-vector variants \cite{xvectors,E-TDNN,dtdnn,lee2020,desplanques20,garciaromero20_odyssey} and xi-vector \cite{lee2021}. All sub-systems developed in I4U adhere to the same development and validation sets as shown in Table \ref{table:dataset_i4u}. We created a common customized development set (I4U dev) comprising of sre18-dev and sre16-eval (30\% trials) since there is no specified dev set in SRE'20. Most parts of the training set were provided by NIST and LDC as listed in Table \ref{table:dataset}. Given the open training condition in SRE'20, each site uses a slightly different training set, with additional data from various sources, which we describe in the following sections. 

\begin{table}[ht!]
\caption{{Common data partitioning across all sites.}}
\begin{center}
    \begin{tabular}{l|l}
  	\toprule
  	Partitions						    & Dataset  				        \\
  	\midrule 
  	Development set				        & (a) sre18-dev, sre16-eval ($30\%$) \\
  	Evaluation set		                & (b) sre20\_cts\_challenge 	\\ 
  	Train set	                        & All datasets except (a) and (b) \\
    \bottomrule
  	\end{tabular}
\end{center}
\label{table:dataset_i4u}
\end{table} 

  \begin{table}
  \caption{{Train set used for open training condition.}}
  \begin{center}
  		\begin{tabular}{l|l}
  			\toprule
  			Corpora							    & Source  			    \\
  			\midrule 
  		    sre16-eval ($70\%$)		            & LDC, NIST             \\
  			SRE'18-eval                         &                       \\
  			SRE'19-eval                         &                       \\
  			SRE'04, 05, 06, 08, 10, 12          &                       \\
  			Switchboard-2 Phase I \& II \& III	& 				        \\
		    Switchboard Cellular Part 1 \& 2	&					    \\
		    Mixer6                              &                       \\
		    Fisher 1 \& 2 				        &  						\\
		    \midrule
  			VoxCeleb 1 \& 2			            & Open         		    \\
  			Private video data set              &                       \\
            \bottomrule
  		\end{tabular}
  	\end{center}
  	\label{table:dataset}
  \end{table} 


\section{Feature Extraction}




\begin{table*}[ht]
    \centering
    \caption{Summary of front-end processing components for each subsystem.}    
    \begin{tabular}{c|c|c|c|c|c}
    \toprule
            & I2R/NUS/Vivo  & LIA & Loquendo & TJU & UEF\\
        \midrule
        Feature set & MFCC  & Fbanks & LogMEL & LogMEL  & PNCC \\
        \# Features & 23 & 60 & 46 & 80 & 30\\
        \# Filters  & 23 & 60 & 46 & 80 & 30\\
        Framing     & 25ms/10ms & 25ms/10ms & 25ms/10ms  & 25ms/10ms & 25ms/10ms \\
        Window      & Hamming & Hamming & Hamming & Hamming & Hamming \\
        SAD         & Energy  & Energy & DNN & Energy & Energy \\
        NFFT        & 512 & 512 & 256 & 400 & 512 \\ 
        Postprocessing & x & Mean & Short-Time Centering  & Mean \& Variation Norm  &x \\ 
    \bottomrule
   \end{tabular}
    \label{tab:front-ends}
\end{table*}

A summary of feature extractors of all subsystems is displayed in Table \ref{tab:front-ends}. We describe them in per-system fashion.

\textbf{I2R/NUS/Vivolab}: We consider mel frequency cepstral coefficient (MFCC) as feature for developing our system. For every utterance, 23-dimensional MFCC features are extracted considering short-term processing by using Hamming windowed frame of 25~ms with a shift of 10~ms. The feature extraction involves 23 mel filetbanks. In addition, the frames with sufficient voice activity are retained by performing an energy-based VAD. 

\textbf{LIA}: For every utterances, 60-dimensional Filter-Banks (FBanks) features are extracted considering short-term processing by using Hamming windowed frame of 25~ms with a shift of 10~ms. A cepstral mean normalization is applied with a window size of 3 seconds. In order to removes silence and low energy speech segments, we used an energy-based VAD.

\textbf{Loquendo and Politecnico di Torino}: We extracted a single set of features for training all the Loquendo DNN embedding systems employed in this evaluation: 46 Log Mel Bands parameters with short time centering (STC) computed on both speech and non-speech audio frames. We used a VAD based on neural network (NN) phonetic decoding. The decoders are hybrid HMM-NN models trained to recognize 11 phone classes. The NN used for the VAD is a multi-layer perceptron that estimates the posterior probability of phonetic units, given an acoustic feature vector. It has been trained on several Speechdat corpora related to different European languages (e.g., English UK, French, German, Italian, Spanish, etc.) using approximately 600 hours of speech.

\textbf{TJU}: We consider log mel filterbanks (LogMEL) as feature for developing our system. For every utterance, 80-dimensional LogMEL features are extracted considering short-term processing by using Hamming windowed frame of 25~ms with a shift of 10~ms. Besides, sentence-level mean and variation normalization is emloyed for feature postprocessing.

\textbf{UEF}: 30-dimensional \emph{Power-normalized cepstral coefficients} (PNCCs) are employed. Our feature extractor differs from the original work~\cite{pncc} by several means: 1) We use mel filterbanks \cite{optimized_pncc} instead of Gammatone filterbanks, with number of triangular filters same as feature dimension. We do not observe notable difference on individual system performance in pilot experiments between the two; 2) We do not perform \emph{discrete cosine transform} (DCT) as post-processing step on the output of power-law nonlinarity. Other operations and related parameters remain same as original work in~\cite{pncc}.

\section{Embeddings and Classifiers}

\subsection{X-vector Neural embedding}
\subsubsection{Extended TDNN}
\label{sec:etdnn}
The Extended-TDNN \cite{E-TDNN} neural network is an evolution of the classic x-vector network \cite{xvectors}, increasing its depth along the frame-level processing block by alternating TDNN layers with feedforward ones. This network is trained according to two different losses: the traditional cross-entropy as well as \emph{additive angular margin loss}~\cite{AAMLoss}.

\subsubsection{Densely Connected TDNN}
\label{sec:dtdnn}
\emph{Densely connected time-delay neural network} (D-TDNN) has been proposed in~\cite{dtdnn}. Its basic unit layer differs from the vanilla TDNN layer by cascading a fully-connected DNN and a TDNN with context expansion. Each of the layers are led by a ReLU activation. The network includes aggregated connection for multiple D-TDNN layers. Our implementation differs from the original network by two aspects: 1) We use attentive statistics pooling (ASP) instead of high-order statistics pooling; 2) When training the network we use additive angular margin softmax function, instead of cosine-based softmax.

\subsubsection{ECAPA TDNN}
The ECAPA TDNN architecture has been originally presented in~\cite{desplanques20}. We observed that the architecture is less accurate for telephonic speaker recognition than the other technological solutions considered, even if it is still moderately helpful for embeddings fusion. The main positive aspect is the fast training and inference time. 

\subsubsection{NEC 43-LAYERS RESNET}
The NEC 43-LAYERS RESNET architecture has been proposed in \cite{lee2020}. This kind of DNN model is very deep, it includes many residuals connections, and it is accurate for telephonic speaker identification. We considered a few implementations with variations in the training and the DNN structure. The most notable solution exploits 53 instead of 43 layers.

\subsubsection{JHU MAGNETO 34-LAYERS RESNET}
The JHU MAGNETO 34-LAYERS RESNET topology is described in \cite{Garcia-Romero2020}. It is based on modified 2D-CNN ResNet-34 architecture. It is by far the slower solution both for training and run-time inferencing because of the presence of 2D Convolutions, whereas all the other considered topologies use 1D Convolution. However, this DNN can achieve extremely good accuracy, in particular for short-duration conditions. We trained many different variants for this architecture to fully exploit its potential. In general, we observed some variability in terms of accuracy, so that the weight initialization and the training hyperparameters seem to be important concerns.

\subsubsection{FTDNN RESNET}
The FTDNN RESNET is a deep ResNet architecture exploiting semi-orthogonal 'factorized' TDNN layers as proposed in \cite{povey18_interspeech}. This provides very good accuracy results for the SRE20 CTS challenge, and it is also much faster than the JHU MAGNETO 34-LAYERS RESNET, in terms of training and run-time computation.

\subsection{Xi-vector Neural embedding}
Xi-vector is the Bayesian counterpart of the x-vector, taking into account the uncertainty estimate. X-vectors, and similar forms of deep speaker embedding, do not consider the uncertainty of features. In a restricted sense, uncertainty is merely captured implicitly with empirical variance estimates at the utterance level. Consequently, they show low robustness against local and random perturbation which is the inherent property of speech utterances. On the technology front, xi-vector offer a simple and straightforward extension to x-vector. It consists of an auxiliary neural net predicting the frame-wise uncertainty of the input sequence. On the theoretical front, xi-vector integrates the Bayesian formulation of linear Gaussian model to speaker-embedding neural networks via the pooling layer. In one sense, xi-vector integrates the Bayesian formulation of the i-vector to that of the x-vector. Hence, we refer to the embedding as the xi-vector, which is pronounced as /zai/ vector. We refer interested readers to \cite{lee2021}.


\section{Sub-system Description}
Listed below are the descriptions of the component classifiers used for I4U fusion. Neural embeddings were used in all sub-systems, with substantial enhancement in terms of network architecture and loss function.


\subsection{I2R}
\subsubsection{Front-end}
We used xi-vector embedding with a simple five-layer TDNN. The training set consists primarily of English speech corpora, which encompasses Switchboard, Fisher, and the MIXER corpora used in SREs 04 -- 06, 08, and 10. We used 23-dimensional MFCCs with 10ms frameshift. Mean-normalization over a sliding window of 3s and energy-based VAD were then applied. Data augmentation \cite{tomko2017} was performed on the training sets using the MUSAN dataset \cite{musan}. 

\subsubsection{Back-end}
Probabilistic linear discriminant analysis (PLDA) was used as the back-end. As in most state-of-the-art implementations, speaker embeddings were reduced to 200 dimensions via \emph{linear discriminant analysis} (LDA) projection before PLDA. 

\subsection{LIA}

\subsubsection{Additional Training Corpora}

The LIA Training corpora used for training the DNN models is composed of I4U corpora and Multilingual LibriSpeech (MLS) datasets~\cite{Pratap2020MLSAL}.

\subsubsection{Front-end}

The x-vector extractor used in the SRE20 CTS challenge is based on wide version of the ResNet-34, that employs a modified version of Squeeze-and-Excitation (SE). The extractor was trained on LIA Training corpora, cut into 4-second chunks and augmented with noise, as described in~\cite{xvectors} and available as a part of the Kaldi-recipe. The speaker embeddings are 256-dimensional and the loss is the angular additive margin with scale equal to 30 and margin equal to 0.2. The size of the feature maps are 256, 256, 512 and 512 for the 4 ResNet blocks. The SE layers are added to the first 2 ResNet blocks. We use stochastic gradient descent with momentum equal to 0.9, a weight decay equal to 2.10$^-4$ and initial learning rate equal to 0.2. The batch size was set to 128, however, training on 4 GPUs in parallel. The implementation is based on PyTorch and the model traning takes about 4 days.

\subsubsection{Back-end}

The back-end of the single system is based on cosine similarity measurement. We obtained on the CTS progress set : EER of 2.83\%, minDCF of 0.129 and actDCF of 0.126.

\subsection{Loquendo and Politecnico di Torino}

\subsubsection{Additional Training Corpora}
Besides the common I4U corpora we also used the following datasets for training the DNN embedding models and the related classifiers:
\begin{itemize}
\item Appen Conversational Telephonic Corpora: 1828 speakers - languages: Bulgarian, Dutch, Hebrew, Croatian, Italian, Portuguese, Romanian, Russian, Turkish  
\item CN-Celeb: 352 speakers - language: Chinese Mandarin
\item Fisher Spanish:  109 speakers – language: Spanish
\item Multi-Language Conversational Telephone Speech 2011: 68 speakers - languages:  American English, Arabic, Czech, Bengali, Dari, Hindi, Persian, Pashto, Slovak, Spanish, South-Asian English, Tamil, Turkish, Urdu
\item Rusten:121 speakers - language: Russian
\end{itemize}

\subsubsection{Front-end: DNN embedding systems}
For the SRE20 CTS challenge, we leveragred large DNN embeddings with more than 20 million parameters to achieve the best possible accuracy. We used PyTorch and we trained eleven DNN embeddings related to the ECAPA TDNN, NEC 43-LAYERS RESNET, JHU MAGNETO 34-LAYERS RESNET, and FTDNN RESNET families, introducing variation in terms of the layers size, the training hyperparameters, and the AMSoftmax loss margin. Table \ref{tab:loq-main-dnn-embeddings}  shows the four standalone DNN embeddings, whose scores have been provided to the I4U consortium for the final fusion.

\begin{table}[ht]
    \centering
    \caption{Main Loquendo DNN embedding systems}
    \begin{tabular}{ c|c|c }
        \toprule
        DNN model & DNN type & Size (millions)\\
        \midrule
        LoqDNN6 & JHU Magneto & 21.1 \\ 
        LoqDNN9 & FTDNN RESNET & 23.3 \\ 
        LoqDNN12 & FTDNN RESNET & 54.7 \\
        LoqDNN18 & JHU Magneto & 28.4 \\
        \bottomrule
        
    \end{tabular}
    \label{tab:loq-main-dnn-embeddings}
\end{table}

Moreover, Table \ref{tab:loq-additional-dnn-embeddings} includes the additional seven DNN embeddings systems that have been used in the embedding fusion systems, described in the next section.

\begin{table}[ht]
    \centering
    \caption{Additional DNNs used in embedding fusion systems}
    \begin{tabular}{ c|c|c }
        \toprule
        DNN model & DNN type & Size(millions)\\
        \midrule
        LoqDNN1 & JHU Magneto & 19.3 \\ 
        LoqDNN2 & ECAPA TDNN & 23.3 \\ 
        LoqDNN3 & NEC 43-RESNET & 24.1 \\
        LoqDNN4 & JHU Magneto & 19.3 \\
        LoqDNN5 & JHU Magneto & 19.8 \\
        LoqDNN7 & NEC 53-RESNET & 32.5 \\
        LoqDNN8 & FTDNN RESNET & 23.3 \\
        \bottomrule
        
    \end{tabular}
    \label{tab:loq-additional-dnn-embeddings}
\end{table}

\subsubsection{Back-end}

The embeddings related to the standalone Loquendo DNNs were transformed in sequence by whitening, LDA projection that reduces their dimensions to 200 and length normalization. The LDA for the Loquendo embedding fusion systems has been computed on all the speakers of the training set used by Loquendo.

The classifier that has been used for obtaining the scores is the \emph{neural pairwise support vector machine} (NeuralPSVM). The classifier takes inspiration from a work carried out by people working in the LEAP lab of the Indian Institute of Science, Bengaluru \cite{ramoji2020pairwise}, \cite{ramoji2020nplda}. The main idea is to perform a neural refinement to pre-trained LDA, whitening, WCCN, and PSVM \cite{cumani2014_1}, \cite{cumani2014_2} scoring matrices. Starting from the initial model, gradient descent iterations are performed by using a smoothed version of NIST DCF as an objective loss function, with two working points: $P_\mathrm{target}$ = 1\% and $P_\mathrm{target}$ = 0.5\%.

Based on the best practices for discriminative PSVM training, impostor pairs were selected among similar speakers. A similar-speaker score matrix was computed offline by using an available accurate text-independent speaker recognition model. The ratio of same-speaker / different-speaker pairs for training was set 16 / 240: this means that, for a given speaker, 16 pairs of utterances of the target speaker are randomly selected and included in the training list, along with 240 pairs of utterances of the selected speaker coupled with randomly selected utterances of the 240 most similar speakers. The batch size was set to a high value (40*4096–68*4096) to have enough impostor pairs for a reliable estimation of the cost function. 

To improve the accuracy, the duration information is added to the embeddings used for the training. The NeuralPSVM classifier includes three additional duration factors related to the sum, the difference, and the minimum number of frames used for extracting the embeddings to compare. The weights of the duration factors are automatically learned by the neural refinement. 

While the standard PSVM is trained on a subset (\textasciitilde 8000) of speakers across the available training corpora the NeuralPSVM classifier is trained on the full training sets including \textasciitilde 14.7 thousand speakers and \textasciitilde 1.8 million utterances, including the original and augmented segments. For the multi-site embedding fusion, we exploited the standard PSVM classifier trained on the subset of corpora used for the LDA.

All our systems but one do not include any score normalization. Only the Emb-fus-1 system exploits the “Cal-Norm” approach, which computes cohort-based statistics on the scores following the approach defined by the adaptive AS-Norm \cite{cumani11_interspeech}. In this case, the mean and the impostor standard deviation values are weighted with predefined coefficients and subtracted from the raw scores as compensation factors.

Table \ref{tab:loq-backend-info} summarizes the LDA, embedding classifier, and score normalization used for the different systems.

\begin{table}[ht]
    \footnotesize
    \centering
    \caption{Scored systems information}
    \begin{tabular}{ c|c|c|c }
        \toprule
        System & LDA size & Classifier & Normalization\\
        \midrule
        LoqDNN6 & 200 & NeuralPSVM & none \\
        LoqDNN9 & 200 & NeuralPSVM & none \\
        LoqDNN12 & 200 & NeuralPSVM & none \\
        LoqDNN18 & 200 & NeuralPSVM & none \\
        Emb-fus-1 & 400 & NeuralPSVM & cal-norm \\ 
        Emb-fus-2 & 450 & NeuralPSVM & none \\ 
        Emb-fus-3 & 450 & PSVM & none \\
        \bottomrule
        
    \end{tabular}
    \label{tab:loq-backend-info}
\end{table}

\subsubsection{Progress Set Results}
Table \ref{tab:loq-performance} summarizes the results of the seven systems provided to I4U consortium on the official SRE20 CTS progress set. The scores of the systems do not include any calibration step for transforming the scores into LLRs at this point.

\begin{table}[ht]
    \centering
    \caption{Performance on the SRE 2020 CTS Progress Set}
    \begin{tabular}{ c|c|c }
        \toprule
        System & EER \% & MinDCF\\
        \midrule
        LoqDNN6 & 3.18\% & 0.117 \\
        LoqDNN9 & 3.06\% & 0.099 \\
        LoqDNN12 & 2.94\% & 0.096 \\
        LoqDNN18 & 3.00\% & 0.108 \\
        Emb-fus-1 & 3.30\% & 0.092 \\ 
        Emb-fus-2 & 3.31\% & 0.109 \\ 
        Emb-fus-3 & 3.66\% & 0.112 \\
        \bottomrule
        
    \end{tabular}
    \label{tab:loq-performance}
\end{table}

\subsubsection{Computational And Memory Requirements}
We used proprietary software for processing the audio files, computing the VAD, and saving LogMel Bands STC information on feature files, excluding the non-speech audio signal portions. On an Intel Core i5-6600, 3.3~GHz server, the program that creates the feature files is \textasciitilde 200 times faster than real-time and requires less than \textasciitilde 15 MB of memory for computing the VAD and performing the feature extraction. 

The feature files related to the audio segments to analyze are then processed with a python script invoking PyTorch for computing the DNN embeddings. We used a server including the Quadro P6000 NVIDIA GPU (having Intel(R) Xeon(R) Gold 625 as CPU) and the embedding extraction procedure exploits GPU computation. Table~\ref{tab:loq-cpu-memory} includes the figures related to the \emph{real-time factor} (RTF) between the processed speech duration and the actual computation time needed by the inference tool. Moreover, Table \ref{tab:loq-cpu-memory} reports also the GPU memory as reported by the API “nvidia\_smi.nvmlDeviceGetMemoryInfo” for all the eleven DNNs prepared by our site for the NIST SRE20 CTS challenge. 

\begin{table}[ht]
    \centering
    \caption{DNN Embeddings Memory and RTFs}
    \begin{tabular}{ c|c|c }
        \toprule
        DNN model & GPU memory & RTF\\
        \midrule
        LoqDNN1 & 19.31 GB & 625x \\ 
        LoqDNN2 & 4.38 GB & 2011x \\ 
        LoqDNN3 & 5.25 GB & 1318x \\
        LoqDNN4 & 18.72 GB & 700x \\
        LoqDNN5 & 19.65 GB & 647x \\
        LoqDNN6 & 20.13 GB & 608x \\
        LoqDNN7 & 6.70 GB & 990x \\
        LoqDNN8 & 5.54 GB & 1274x \\
        LoqDNN9 & 5.54 GB & 1271x \\
        LoqDNN12 & 20.2 GB & 565x \\
        LoqDNN18 & 6.25 GB & 789x \\
        \bottomrule
    \end{tabular}
    \label{tab:loq-cpu-memory}
\end{table}

Finally, the scoring for this evaluation has been performed by using a non-optimized Python environment. The elapsed time for obtaining the raw PSVM scores for the \textasciitilde 2.6 M trials is around a couple of minutes. Obtaining the Cal-Norm scores requires about less than 3 hours. It is worth noting, however, that this implementation has been used just for these experiments. It has been designed to favor flexibility over computational efficiency.

\subsection{NEC}
\subsubsection{Front-end}
We used a variant of x-vector extractors which had a 43 TDNN layers with residual connections. This is exactly same as the one shown in the work~\cite{lee2020}. Here, 2-head attentive statistics pooling was used in the same way as in the paper. Additive margin softmax loss was used for optimization. 512-dimension bottleneck features from the first segment-level layer was used as speaker embeddings.

\subsubsection{Back-end}
Heavy-tailed PLDA (HT-PLDA) was used as the back-end in our systems. NIST SRE 04-12 datasets were used for producing out-of-domain PLDA. On the other hand, we also trained in-domain PLDA using SRE16 and SRE18 set.
X-vectors for in-domain PLDA training were centerized using SRE16, SRE18 and SRE19 set. Then we applied linear interpolation between the out-of-domain PLDA and the in-domain PLDA. Weights for both PLDAs were 0.5.

\subsection{NUS}
\subsubsection{Front-end}
We perform data augmentation on the training dataset, which comprise of previous editions of SRE datasets 2004-2016, Mixer6, Switchboard, Fisher and VoxCeleb1-2 datasets. The number of training utterances are doubled by adding noisy and reverberated versions of the clean utterances. For this purpose, music, speech and babble noise segments extracted from the MUSAN database is used~\cite{musan}. The MFCC features of the augmented training set are extracted for training a TDNN model based on the architecture described in~\cite{xvectors} to extract 512-dimensional embeddings.

\subsubsection{Back-end}
The back-end of the system considers a 150-dimensional LDA to reduce the dimension of the x-vectors followed by PLDA classifier to compute the likelihood scores. It is noted that the PLDA model is first adapted with SRE 2018 eval set before scoring, which we found to give a better result than that without domain adaptation. Additionally, score normalization is applied with adaptive s-norm technique (30\% top scores) considering the SRE 2018 eval set.

\subsection{THUEE}
\subsubsection{Front-end}
Training data we used includes SRE04-10, MIXER6, Switchboard (SWBD), Voxceleb 1\&2 and Fisher datasets. These datasets (i.e., SRE, SWBD, Voxceleb) are augmented by different folds for different systems after convolving with far-field Room Impulse Responses (RIRs), or by adding noise from the MUSAN corpus. MFCC as acoustic feature is extracted with a 25~ms window size and a time shift of 10~ms. The operations of data augmentation and feature extraction are dependent on Kaldi x-vector recipe. We used the extended factorized TDNN (EF-TDNN) as baseline model and did some extensions on it called EF-TDNN\_LSTMP and EF-TDNN\_SE respectively. EF-TDNN\_LSTMP, compared with basic EF-TDNN, replace the 19th layer with \emph{long
short-term memory with recurrent project layer} (LSTMP)~\cite{LSTMP}. LSTMP is combination of two improved methods. One is to introduce a recurrent projection layer between the LSTM layer and the output layer. The other is to introduce another non-recurrent projection layer to increase the projection layer size without adding more recurrent connections. In our LSTMP layer, it has 1024-dimensional cell, 512 recurrent and non-recurrent projection dimensions. Besides, considering the channel attention mechanism from the squeeze-and-excitation (SE) block \cite{SqueezeandExcitation} and its previous performance on different fields, EF-TDNN\_SE mainly changes the structure of F-TDNN layer by adding SE layer with reduction ratio as 16 before the last convolutional layer.
\subsubsection{Back-end}
After the embeddings are extracted, they are then transformed to 300 dimension using LDA. Then, embeddings are projected into
unit sphere. At last, adapted PLDA with no dimension reduction is applied. As for the data for LDA/PLDA adaptation, it is achieved by filtering more adaptive datasets according to the results on the leader board of CTS chanllenge.

\subsection{TJU}
\subsubsection{Front-end}
We use the SpeechBrain \cite{speechbrain} toolkit to realize a standard ECAPA-TDNN system. After removing erroneous labels, 7,447 speakers are finally selected from SRE datasets 2004-2016 with 120,443 utterances. The network inputs 80-dimensional log fbank features and ouputs 192-dimensional speaker embeddings. Time domain SpecAugmentation, adding noise, adding reverberation, and adding noise \& reverberation at the same time are implemented for data augmentation. Speeches are cropped and reunited after Voice Activity Detection (VAD). The whole process of features extraction are performed with GPUs using on-the-fly approach. Sentence-level normalization is used for input features. Additive Angular Margin Softmax and CyclicLRScheduler are employed to train the system.
\subsubsection{Back-end}
The back-end of the single system is based on cosine similarity measurement, with EER of 4.01\%, MinDCF of 0.247, and ActDCF of 0.512 in the CTS eval set.

\subsection{UEF}
\subsubsection{Front-end}
Our speaker embedding extractor is based on the D-TDNN described in section \ref{sec:dtdnn}. Our training data is composed of VoxCeleb1 training set, VoxCeleb2, LibriSpeech, Switchboard, MIXER-6, and SRE datasets (2004-2010). After removing very short utterances (less than 4~s) and speakers with too few number of utterances (less than 8), the total number of speakers are 12860. We perform data augmentation using \emph{room impulse response} (RIR) \cite{tomko2017} and MUSAN dataset \cite{musan}, followed by random sampling a subset which is two times larger than the original data. We combined the subset and the original set as the training data for D-TDNN. Adam \cite{adam} is adopted as the optimizer. Speaker embeddings are extracted from the first fully-connected layer after pooling.

\subsubsection{Back-end}
The extracted speaker embeddings are mean subtracted and length normalized, before being transformed to 200 dimension using LDA. We used SRE2004-2010, Switchboard and MIXER-6 to train PLDA, and SRE20 development set for adapting it~\cite{kaldi_plda}. Log-likelihood scores provided by this sub-system are returned by the adapted PLDA.

\subsection{Vivolab and Agnitio}
\subsubsection{Front-end}
Vivolab embedding extractor is built based on the Extended-TDNN described in section \ref{sec:etdnn}. The training corpus consists of the traditional MIXER6 corpora (SRE04-06, SRE08 and SRE10), complemented with the data from SRE18 and VoxCeleb 1. Agnitio embedding extractor is also based on Extended-TDNN neural network trained with the same data as Loquendo and Politecnico di Torino do. As input features, Agnitio selected wav2vec \cite{wav2vec}.

\subsubsection{Back-end}
Vivolab backend is based on the state-of-the-art LDA-PLDA pipeline. For this purpose we define two subsets of data to take into consideration: On the one hand we include both MIXER6 and VoxCeleb1 data as a large out-of-domain subset. On the other hand the in-domain subsets consists of excerpts from SRE16, SRE18eval and SRE19. Due to the different nature between both subsets we first apply CORAL \cite{CORAL}, weighting both domains with a factor of 0.5. The adapted embedings now undergo centering and LDA-based whitening (a reduction of dimension up to 300). Final scores take into account two different models,  the simplified PLDA (SPLDA) as well as a joint PLDA of two factors \cite{tiedPLDA}, with a inter-speaker subspace dimension of 100. These models consider the same training corpus as CORAL and the LDA. Agnitio backend consists on LDA-PSVM trained with the same data as Loquendo and Politecnico di Torino do.

\section{Fusion and Calibration}

\subsection{Embedding Fusion}
To enhance the accuracy, we also exploited embedding fusion approaches by stacking x-vectors produced by different systems. This allows exploiting the orthogonality of different systems and obtaining better accuracy results. It is worth remarking that such an approach is computationally expensive, and it makes sense mainly on challenges and evaluations to showcase technology.

Currently, we provided scores to the I4U consortium related to three embedding fusions. Two of them were based on the fusion of Loquendo DNN embeddings systems. The remaining DNN embedding fusion involved using embeddings produced by multiple I4U sites. Table \ref{tab:loq-embedding-fusions} summarizes the embedding fusion systems used in the challenge and their stacked size. 

Similar to the embeddings related to the standalone Loquendo DNNs, fused embedding were transformed in sequence by whitening, LDA projection that reduces their dimensions (to 400 or 450) and length normalization. Moreover, the LDA for the Loquendo embedding fusion systems has been computed on all the speakers of the training set used by Loquendo, while for the multi-site embedding fusion the LDA has been computed on a common subset of corpora including NIST SRE04-10, Mixer6, SRE16, SRE18, and SRE19. Tables \ref{tab:loq-backend-info} and \ref{tab:loq-performance} compared the fused embeddings with that of standalone embeddings.

\begin{table}[ht]
    \footnotesize
    \caption{Embedding fusion systems}
    \centering
    \begin{tabular}{ c|c|c }
        \toprule
        Model Name & DNN models & Stacked size\\
        \midrule
        Emb-fus-1 & LoqDNN1-2-3-5 & 2048 \\ 
        Emb-fus-2 & LoqDNN4-6-7-8-9 & 2304 \\ 
        Emb-fus-3 & I2R,LOQDNN6-9, NEC,TJU,UEF & 3072 \\
        \bottomrule
    \end{tabular}
    \label{tab:loq-embedding-fusions}
\end{table}

\subsection{Score Fusion and Calibration}
For score fusion, a Python implementation\footnote{\url{https://gitlab.eurecom.fr/nautsch/pybosaris}} of the BOSARIS toolkit \cite{Brummer-deVilliers-BOSARIS-Binary-Scores-AGNITIO-Research-2011} was used for logistic regression yielding  log-likelihood ratio (LLR) scores. In preliminary studies, we \emph{(i)} investigated and extended quality estimates \cite{Ferrer-UAC-2012,miranti-btfs-2013}, and \emph{(ii)} observed that some systems are causing worse performance if included (especially on the progress set). 

We found log-duration to be helpful despite data shifts. In Miranti's method \cite{miranti-btfs-2013}, log-ratios of both audio segments' durations are investigated. We generalise and let the regression find exponents for each duration value. Of $n$ systems, scores $S_i$ are fused to $S'$ using $n+3$ weights $w_0, w_i, w_r, w_p$ and depending durations of reference (summed if multiple files) and probe audios $d_r, d_p$ (also easier to implement):
\begin{align}
    S' = w_0 + \sum_{i \in 1..n} w_i\,S_i + w_r\,\log(d_r) + w_p\,\log(d_p).
\end{align}

When fusing all above described subsystems and embeddings fusion systems, we investigated the individual LLR contribution of each subsystem ($w_i\,S_i$). Looking at the absolute value of a system's highest/lowest contribution, we observed that some contributed small offsets of $\leq 1$ to any LLR; systems of weak contribution. Jackknifing showed that some remained useful nonetheless. Our fusion comprised twelve subsystems and the two duration sets.\footnote{Since the challenge rules were that systems of lower actual DCF will become the new primary system, we added a $-2$ offset to all our scores, so we could update potentially later with lower minDCF systems.}

\begin{table}[ht]
    \centering
    \caption{Initial analysis: LLR contribution potentials.}
    \label{tab:llr-contribution-pre-assessment}
    \begin{tabular}{c|c|c|c}
        \toprule
        Subsystem & $\min w_i\,S_i$ & $\max w_i\,S_i$ & Used for fusion \\
        \midrule
        Emd-fus-1 & -6.3 & 2.6 & x \\
        Emb-fus-2 & -9.0 & 5.7 & x  \\
        Emb-fus-3 & -6.8 & 2.6 & x  \\
        \midrule
        I2R & -0.2 & 0.1 \\
        I2Rr1 & -2.6 & 1.1 \\
        I2Rr2 & -0.7 & 1.7 & x  \\
        I2Rr3 & -0.7 & 1.9 \\
        I2Rr4 & -8.1 & 3.5 & x  \\
        \midrule
        LoqDNN6 & -0.9 & 1.7 & x  \\
        LoqDNN9 & -0.1 & 0.1 \\
        LoqDNN12 & -3.9 & 2.2 & x  \\
        LoqDNN18 & -0.9 & 1.6 \\
        \midrule
        LIA & -0.8 & 0.4 & x  \\
        NEC & -4.8 & 2.5 & x  \\
        THUEE & -2.6 & 1.3 & x  \\
        UEF & -0.4 & 0.7 & x \\
        VIVO & -1.8 & 0.5 & x  \\
        \bottomrule
    \end{tabular}
\end{table}

Table~\ref{tab:llr-contribution-pre-assessment} shows the contribution to fusion of the above subsystems in a preliminary assessment fusion. Neither are durations included, nor is the decision for inclusion in the final fusion set-up made solely on this information. Yet, for this composition, one can identify the embedding fusions to be the draught horses here. While individual systems can perform good, their contribution here might be little when they are many alike systems in that particular composition (weights depend on the composition of systems to be fused); vice versa, systems of low individual performance might appear better here for they behave differently. Moreover, some systems of little contribution have critical impact on the small yet remaining performance gains on progress set.

The contribution to the I4U fusion is shown in Table~\ref{tab:llr-contribution-fusion}. Weights are shown for curtosy reasons; some are negative---some systems became correctors to an otherwise overconfident fusion outcome (if they would not have been a part of the fusion). Log-durations are demonstrated for their need as a correction measure; as a quality estimate. Among the subsystems, the ranges of LLR contribution changed slightly. In comparison to the subsystems' contribution to the fusion LLR, log-duration appears not considerable to yield an LLR by itself; yet, it re-adjusts for too high LLR estimates. There are two impacts of weights: proportion a subsystem matters within the fusion composition while also making its scores fit to the LLR scale of composed systems.

\begin{table}[ht]
    \centering
    \caption{Fusion: LLR contributions ($w_0 = 12.5$).}
    \label{tab:llr-contribution-fusion}
    \begin{tabular}{c|c|c|c}
        \toprule
        Subsystem & $\min w_i\,S_i$ & $\max w_i\,S_i$ & weight $\times100$\\
        \midrule
        Emd-fus-1 & -6.1 & 2.5 & 27.6 \\
        Emb-fus-2 & -8.3 & 5.2 & 50.8 \\
        Emb-fus-3 & -6.8 & 2.6 & 55.5 \\
        \midrule
        I2Rr2 & -0.7 & 1.8 & -3.5 \\
        I2Rr4 & -7.2 & 3.1 & 13.5 \\
        \midrule
        LoqDNN6 & -0.7 & 1.5 & -13.1 \\
        LoqDNN12 & -3.6 & 1.0 & 25.1 \\
        \midrule
        LIA & -0.9 & 0.4 & 0.5 \\
        NEC & -4.1 & 2.2 & 2.7 \\
        THUEE & -2.9 & 1.5 & 1.0 \\
        UEF & -0.5 & 0.9 & -0.6 \\
        VIVO & -0.2 & 0.0 & -0.8 \\
        \midrule
        $\log(d_r)$ & -4.0 & -3.3 & -39.8 \\
        $\log(d_p)$ & -4.7 & -3.2 & -52.7 \\
        \bottomrule
    \end{tabular}
\end{table}

\section{Results on Development and Progress Sets}
\label{sec:results}
Performance of the submissions on I4U Development set (as in Table \ref{table:dataset_i4u}) and Progress (sre20\_progress) Sets are shown in Table \ref{table:performance_sre19dev}.

 \begin{table} [ht!]
  	\caption{{Performance of the primary fusion on I4U Development and SRE'20 Evaluation Sets.}}
  	\begin{center}
  	   \begin{tabular}{l|c|c|c}
  			\toprule
			\textbf{Equalized}  & EER (\%) 	& Min $C_{primary}$ & Act $C_{primary}$	\\ 
			\midrule
  			I4U Dev 		& 2.18  & 	0.038   & 	0.043 \\
   			Progress	& 2.53  & 	0.077	& 	0.094 \\
   			Test        & 2.91  & 	0.066   & 	0.070 \\
  			\bottomrule
    \end{tabular}      
    \end{center}
  	\label{table:performance_sre19dev}
 \end{table}

\section{Computation and Memory Requirement}
The GPU and CPU time for individual sub-systems to process a single trial is shown in Table~\ref{table:cpu_time}. Peak memory usage required to process a pair of enrollment and test recordings was approximately $300$ to $700$ MB. 

\begin{table}[h!]
  \small
  \caption{{CPU and GPU execution time used to process a single trial for each sub-system/sub-fusion in term of real-time factor (RT).}}
  \begin{center}
  \begin{tabular}{c|c|c}
  \toprule
	Sub-system      & RT (CPU)  &RT (GPU)       \\ 
	\midrule 
	Emd-fus-1       & -         & 0.0044        \\
	Emd-fus-2       & -         & 0.0056        \\
	Emd-fus-3       & 3.60      & 0.0094        \\
	I2R (r2, r4)    & 0.281     & -             \\
	LoqDNN6         & -         & 0.0015        \\
	LoqDNN12        & -         & 0.0018        \\
	LIA             & -         & 0.0017        \\
 	NUS             & -         & 0.0020        \\
 	NEC             & 0.417     & -             \\
 	THUEE           & -         & -             \\
 	UEF             & 2.902     & 0.0012             \\
 	VIVO            & -         & -             \\
 	\bottomrule
 	\end{tabular}
  \end{center}
  \label{table:cpu_time}
\end{table} 

\section{Conclusion} 

The I4U submissions were based on the score fusion of multiple  sub-systems.   Scores  from  individual  sub-system/sub-fusion were first calibrated followed by simple linear transformation fusion. Another highlight to I4U submission is embedding fusion, where sub-system embeddings are concatenated and used as input to the classifier. 

\section{Acknowledgements} 
This work was partially funded by the ANR/JST VoicePersonae project (grant No. JPMJCR18A6).

\balance
\bibliographystyle{IEEEbib}
\bibliography{ref}



\end{document}